\documentclass[lettersize,journal]{IEEEtran}
\usepackage{amsmath,amsfonts}
\usepackage{algorithmicx}
\usepackage{algorithm}
\usepackage{array}
\usepackage{textcomp}
\usepackage{stfloats}
\usepackage{url}
\usepackage{verbatim}
\usepackage{graphicx}
\usepackage{cite}
\usepackage{svg}
\usepackage{booktabs, threeparttable, caption, makecell, subcaption}
\usepackage{multirow}
\usepackage{adjustbox}
\usepackage{algpseudocode}

\DeclareMathOperator*{\argmin}{argmin}

\begin{document}

\title{Memory-Efficient Training for Deep Speaker Embedding Learning in Speaker Verification}

\author{Bei Liu \IEEEmembership{Student Member, IEEE,}
        and Yanmin Qian, \IEEEmembership{Senior Member, IEEE}
\thanks{
Both the authors are with the Auditory Cognition and Computational Acoustics Lab, the Department of Computer Science and Engineering and the MoE Key Laboratory of Artifificial Intelligence, AI Institute, Shanghai Jiao Tong University, Shanghai 200240, China (e-mail:\{beiliu, yanminqian\}@sjtu.edu.cn)}

}

\markboth{Journal of \LaTeX\ Class Files,~Vol.~14, No.~8, August~2021}%
{Shell \MakeLowercase{\textit{et al.}}: A Sample Article Using IEEEtran.cls for IEEE Journals}


\maketitle

\begin{abstract}
Recent speaker verification (SV) systems have shown a trend toward adopting deeper speaker embedding extractors. Although deeper and larger neural networks can significantly improve performance, their substantial memory requirements hinder training on consumer GPUs. In this paper, we explore a memory-efficient training strategy for deep speaker embedding learning in resource-constrained scenarios. Firstly, we conduct a systematic analysis of GPU memory allocation during SV system training. Empirical observations show that activations and optimizer states are the main sources of memory consumption. For activations, we design two types of reversible neural networks which eliminate the need to store intermediate activations during back-propagation, thereby significantly reducing memory usage without performance loss. For optimizer states, we introduce a dynamic quantization approach that replaces the original 32-bit floating-point values with a dynamic tree-based 8-bit data type. Experimental results on VoxCeleb demonstrate that the reversible variants of ResNets and DF-ResNets can perform training without the need to cache activations in GPU memory. In addition, the 8-bit versions of SGD and Adam save 75\% of memory costs while maintaining performance compared to their 32-bit counterparts. Finally, a detailed comparison of memory usage and performance indicates that our proposed models achieve up to 16.2x memory savings, with nearly identical parameters and performance compared to the vanilla systems. In contrast to the previous need for multiple high-end GPUs such as the A100, we can effectively train deep speaker embedding extractors with just one or two consumer-level 2080Ti GPUs.
\end{abstract}


\section{Introduction}
\label{sec:intro}
\IEEEPARstart{S}{peaker} verification (SV) is a crucial biometric identification technology that aims to confirm the identity of a person by examining the distinct features of their voice. A typical SV system comprises two main computational components: a front-end speaker embedding extractor and a back-end similarity calculator to compare embeddings. Before the era of deep learning, i-vector~\cite{ivector} coupled with probabilistic linear discriminant analysis (PLDA)~\cite{plda} provides the most competitive results. In recent decades, neural networks have been widely adopted as speaker embedding extractors, achieving remarkable performance~\cite{tandem}. x-vector~\cite{xvector} presents a standard paradigm which contains three different parts: a stack of layers that process frame-level features, a pooling layer that aggregates segment-level embeddings, and a softmax layer that generates probabilities for each speaker. Firstly, a series of network layers operate on the frame-level acoustic features of a given utterance. Then, a statistical pooling layer extracts a fixed-length speaker embedding. Finally, the entire system is trained with cross-entropy loss. Follow-up works have explored a range of strategies to boost system performance, including advanced network architectures~\cite{tdnn-sv, xvector, ext-xvector, rvector, d-tdnn, ecapa, ecapa++, dense-residual, skd, adaptive-cnn, df-resnet, df-resnet-journal, mlp}, attentive pooling techniques~\cite{pooling1, pooling2, pooling3, pooling4, pooling5}, and loss functions that are aware of hard samples~\cite{sv-loss1, sv-loss2, sv-loss3, sv-loss4}.

In terms of network architectures, a variety of backbones have been put forward in recent years. ~\cite{xvector} initially applies a time delay neural network (TDNN) to the speaker verification task. Subsequent, ECAPA-TDNN~\cite{ecapa} proposes a multi-scale computational block that integrates features from various hierarchical levels. ECAPA++~\cite{ecapa++} further pushes the limits of performance by incorporating a recursive convolution module to capture fine-grained speaker information. Another line of research focuses on convolutional neural networks (CNN). In VoxSRC 2019, Zeinali et al.~\cite{rvector} adopted ResNet~\cite{ResNet} as the embedding extractor and achieved the first place. Since then, multiple lightweight attention mechanisms, such as coordinate attention~\cite{aff}, simple attention~\cite{simple-att} and triplet attention~\cite{dpnet}, have been proposed to improve its performance. In addition, the winner of VoxSRC 2021 employed a re-parameterized model named RepVGG~\cite{repvgg}. Alternatively, ~\cite{df-resnet} and ~\cite{df-resnet-journal} suggest that depth is more important than width, resulting in the introduction of a novel DF-ResNet model with an astonishing 233 layers. In addition, Chen et al. deepened ResNet to 293 layers in CNSRC 2022~\cite{cnsrc-2022} and VoxSRC 2022~\cite{voxsrc-2022}. Recently, the UNISOUND system~\cite{unisound} for VoxSRC 2023 sets a new milestone by increasing ResNet layers to 518.
   
Recent speaker verification challenges have indicated a clear trend for speaker embedding extractors to become progressively deeper. Although deeper and larger neural networks can yield exceptional results, their substantial memory requirements hinder the training of deep speaker embedding extractors on consumer GPUs. For example, the DF-ResNet233 model, as proposed in~\cite{df-resnet} and ~\cite{df-resnet-journal}, is trained using four A100 GPUs, each with 40GB of memory. ~\cite{unisound} utilizes 10 V100 GPUs (32GB) for training ResNet314 and 60 V100 GPUs (32GB) for ResNet518. These computational resources are inaccessible and unaffordable to many researchers. To reduce memory costs in training large-scale SV systems, one possible approach is to reduce the batch size. However, using excessively small batch sizes can impair system performance due to inaccurate gradient estimation and batch normalization, and it can also significantly slow down the entire training process. How to achieve memory-efficient training for deep speaker embedding learning is a crucial and challenging topic in the SV field. This paper investigates the possibility of effectively training very deep networks in resource-limited scenarios for speaker verification.

Firstly, we conduct a systematic examination of GPU memory allocation throughout the training process of speaker verification systems. Modern neural networks are generally trained with back-propagation algorithm~\cite{backprop}. The network weights are first loaded onto the GPU, and for each training batch, the forward pass computes activations. In the backward pass, gradients are calculated to update the weights. To accelerate convergence, commonly used optimizers are stateful, which track gradient statistics over time. For example, stochastic gradient descent (SGD) with momentum~\cite{sgd} accumulates historical gradients for each weight. Adam~\cite{adam} and its variant AdamW~\cite{adamw} maintain two states: gradient and the second-order gradient. In brief, GPU memory usage mainly comes from three sources: weights, activations and optimizer states. While reducing the number of weights can lower memory costs, it may also cause performance degradation, which will not be covered in this work. Instead, we will focus on optimizing GPU memory allocation from the perspectives of activations and optimizer states.

Regarding activations, mainstream deep learning frameworks like TensorFlow~\cite{tensorflow} and PyTorch~\cite{pytorch} store activations during the forward pass for fast gradient computation in the backward pass. Although this approach speeds up training, it causes memory usage to increase linearly with network depth. This involves a trade-off between space and time. The current method sacrifices space for time. However, the limited memory capacity of GPUs makes it difficult to accommodate very deep neural networks~\cite{memory_wall}. To address this issue, we introduce a reversible computational principle, which allows back-propagation without the need to store activations in memory. By applying this principle to ResNets and DF-ResNets, we develop two new families of reversible neural networks: RevNets and DF-RevNets. Specifically, two types of reversible networks, namely partially and fully reversible networks, are presented. They consist of a series of reversible blocks, where the activations of each layer can be recovered from the previous layer during the backward pass. Consequently, both RevNets and DF-RevNets maintain almost constant memory usage, regardless of the network's depth.

Moreover, optimizer states are another source of memory allocation. Generally, gradients and related statistics are stored using 32-bit floating-point numbers, meaning that SGD and Adam will occupy 4 and 8 bytes per weight, respectively. To reduce the memory costs of optimizer states, we present a dynamic quantization scheme that converts 32-bit states into 8-bit. Initially, the optimizer state tensor is partitioned into smaller blocks, each of which is independently quantized to mitigate the outlier effect and reduce quantization errors. These blocks are then normalized to the range $[-1, 1]$ by dividing their absolute maximum values. A dynamic tree-based 8-bit data type is utilized to ensure precision for both small and large magnitudes. Finally, a binary search is conducted to find the nearest 8-bit value to the normalized tensor, and the corresponding integer index is stored. For parameter updates, the optimizer states are first converted from 8-bit to 32-bit. Weight updates are then performed, and the states are quantized back to 8-bit for storage. By integrating this scheme with commonly used optimizers, we create 8-bit versions of SGD and Adam, achieving a 75\% reduction in memory cost while maintaining performance nearly identical to that of 32-bit optimizer states.

To summarize, the key contributions of this study are highlighted below. To the best of our knowledge, this is the first research to thoroughly investigate memory-efficient training methods for deep speaker embedding learning in speaker verification under the resource-limited training scenarios.

\begin{enumerate}
    \item Firstly, we conduct a systematic study on GPU memory allocation during the training of SV systems. Empirical analyses reveal that, aside from weights, activations and optimizer states are the main consumers of memory.
    \item For activations, a reversible computational principle is introduced to perform back-propagation without storing activations in memory. We develop two different types of reversible neural networks that significantly reduce memory costs without performance loss.
    \item For optimizer states, we propose a 8-bit dynamic quantization strategy to replace 32-bit floating-point numbers. The 8-bit versions of SGD and Adam save 75\% of memory usage while maintaining performance.
    \item We evaluate the proposed methods on popular deep speaker embedding extractors, including ResNet101, ResNet152, DF-ResNet110, DF-ResNet179, and DF-ResNet233, showing their effectiveness and generality. 
    \item Finally, a detailed analysis of memory utilization and performance indicates that up to 16.2x memory savings can be achieved with nearly identical parameters and performance compared to the vanilla systems.
\end{enumerate}

This is an extended work of our previous paper~\cite{reverb}. We first systematically examine the training GPU memory allocation for speaker verification systems. Furthermore, a dynamic quantization scheme is presented to compress 32-bit optimizer states into 8-bit without performance loss. During the experiments, we conduct an extended evaluation on a family of DF-ResNets, including DF-ResNet56, DF-ResNet110, DF-ResNet179 and DF-ResNet233. Additionally, a thorough analysis of memory usage and performance is provided, highlighting the effectiveness and broad applicability of our memory-efficient training approaches for various deep speaker embedding extractors.

\section{GPU Memory Allocation Analysis in Speaker Modeling Training Phase}

In this section, we first review the computational process of the back-propagation algorithm used in modern neural network training. Then, we analyze the GPU memory allocation of speaker verification systems based on ResNet and DF-ResNet during the training phase.

\subsection{Back-propagation Algorithm}
\label{ssec:back_prob}
 The back-propagation algorithm~\cite{backprop}, often referred to as reverse mode automatic differentiation, utilizes the chain rule to compute the gradient of neural network parameters with respect to a given loss function by traversing the computational graph in reverse. This algorithm is the standard choice for training modern neural networks and is extensively used in deep learning libraries such as TensorFlow~\cite{tensorflow}, PyTorch~\cite{pytorch}, and JAX~\cite{jax}.

 In a neural network, each layer with trainable weights can be treated as a node in the computational graph during training. The back-propagation algorithm is implemented by constructing this computational graph. For a given computational graph $\mathcal{G}$, the topological ordering of its nodes can be represented as $v_1, \ldots, v_l$, where $v_l$ denotes the loss function $\mathcal{L}$. Each node is characterized by a function $f_i$ that depends on its parent node and contains learnable parameters $w_i$. Let $a_i$ represent the activation value of the $i$-th node (i.e., the output result of the node). The detailed calculation process of the back-propagation algorithm is as follows:

\begin{itemize}
  \item Forward-propagation: As the training data is fed into the neural network, activations are calculated for each node $v_i$ in a forward pass, following the topological order of the computational graph. The activation $a^i$ for each node are stored for later use in gradient computation during the back-propagation phase.
  \begin{align}
    a^{i+1} &= v_i(a^i) \label{eq:backprob1}
  \end{align}
  
  \item Loss calculation: The loss $\mathcal{L}$ is computed based on the neural network's predictions and the true labels, and it is a function of the trainable weights throughout all the layers of the network.
  \item Back-propagation: In this step, the weights $w_i$ are updated by computing the gradient of the loss with respect to each weight, followed by the next iteration of calculations. The back-propagation algorithm applies the chain rule to compute the total derivative $d\mathcal{L} / dv_i$ for each node $v_i$ in reverse topological order.
  \begin{align}
    \frac{d\mathcal{L}}{dv_i} &= \sum_{j \in Child(i)}\biggl(\frac{\partial f_j}{\partial v_i}\biggr)^T \frac{d\mathcal{L}}{dv_j} \label{eq:backprob2}
  \end{align}
  where $Child(i)$ represents the child node of $v_i$ in the computational graph $\mathcal{G}$. $\partial f_j / \partial v_i$ means the Jacobian Matrix, which is used to calculate the partial derivative of $f_j$ with respect to node $v_i$.

  \item Parameter update: The learnable parameters $wi$ are updated using the total derivative $d\mathcal{L} / dv_i$ for each node obtained in the previous step. This process is repeated until the neural network training converges.
  \begin{align}
    w_i \gets w_i - \eta \frac{d\mathcal{L}}{dv_i} \label{eq:backprob3}
  \end{align}
  where $\eta$ is the learning rate.
  
\end{itemize}

\subsection{GPU Memory Allocation Analysis}
\label{ssec:alloc_analysis}
During training, there is a trade-off between storage space and training time. The prevailing approach is to reduce training time at the expense of storage. In forward propagation, the activations of each node are stored for later use in gradient calculations in back-propagation pass, thus accelerating the training process. However, this method increases GPU memory consumption, with memory usage growing linearly as network depth increases. Given the limited GPU memory capacity~\cite{memory_wall}, this restricts the training of very deep neural networks, particularly in resource-constrained scenarios, making it difficult to train large-scale speaker verification systems. In this section, we will examine GPU memory allocation during the training of ResNet- and DF-ResNet-based Speaker Verification systems and investigate strategies to effectively reduce GPU memory usage.

\begin{figure}[t]
  \centering
  \includegraphics[width=\linewidth]{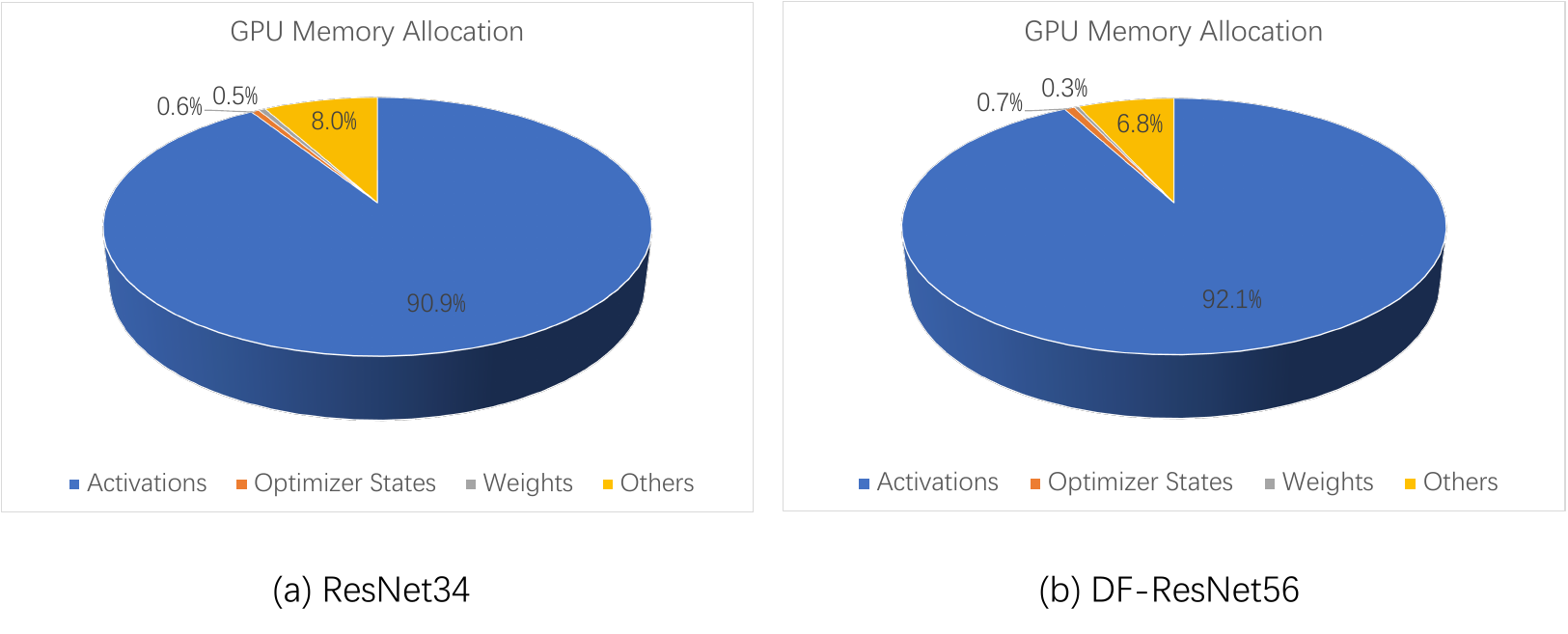}
  \caption{GPU memory allocation during the training process for ResNet34 and DF-ResNet56 in speaker verification.}
  \label{fig:memalloc}
\end{figure}

\begin{table}[t]
  \caption{Detailed GPU memory consumption for ResNet34 and DF-ResNet56 in speaker verification.}
  \label{tab:memalloc}
  \centering
  \begin{adjustbox}{width=.48\textwidth,center}
  \begin{tabular}{c|c|c|c|c}
    \toprule
    \textbf{GPU Memory Allocation} & \textbf{ResNet34} & \% & \textbf{DF-ResNet56} & \%   \\
    \midrule
    Activations & 5.14GB & 90.9 & 5.244GB & 92.1  \\
    \midrule
    Optimizer States & 0.032GB  & 0.6 & 0.041GB & 0.7  \\
    \midrule
    Weights & 0.027GB  & 0.5 & 0.018GB & 0.3  \\
    \midrule
    Others & 0.455GB  & 8.0 & 0.389GB & 6.8  \\
    \bottomrule
  \end{tabular}
  \end{adjustbox}
\end{table}

In the experiments, we use ResNet34 and DF-ResNet56 as examples to demonstrate the details of GPU memory usage. The training was conducted using the PyTorch framework on a 2080Ti GPU with 11GB of memory. The batch size was set to 64 for ResNet34 and 16 for DF-ResNet56. The results are presented in Fig. \ref{fig:memalloc} and Table \ref{tab:memalloc}. We can clearly see that for both ResNet34 and DF-ResNet56, activations account for the highest GPU memory usage. For example, activations make up 90.9\% of the total memory consumption for ResNet34, and 92.1\% for DF-ResNet56. This is consistent with previous analysis: during the backpropagation algorithm, the activations of each node must be stored, requiring corresponding GPU memory allocation. Moreover, as the number of layers in the neural network increases, the amount of cached activations grows, leading to a linear rise in GPU memory usage.

In addition, two other sources of memory usage arise from the weights and optimizer states. At the start of training, the network weights need be loaded onto the GPU. By default, these weights are stored as 32-bit floating-point numbers, consuming a certain amount of memory. For instance, the weights of ResNet34 occupy 0.5\% of the memory, while DF-ResNet56 uses 0.3\%. During back-propagation, the gradients of the loss function with respect to the weights are computed and stored, which are then used to update the weights. Common optimizers, such as SGD and Adam, not only store these gradients but also track historical statistics to enhance training stability and accelerate convergence. For example, Adam maintains second-order gradient information. These gradients and optimizer states are also stored in 32-bit floating-point format. The memory overhead for optimizers is 0.6\% for ResNet34 and 0.7\% for DF-ResNet56.

Besides, the PyTorch framework itself consumes a portion of GPU memory. This overhead is related to the software infrastructure of the deep learning framework and is independent of the neural network architecture and input data. Hence, it is categorized as "others" and is not within the topic of this paper.

From the above analysis, it is clear that the primary sources of GPU memory usage come from activations, optimizer states, and weights. While reducing the number of parameters can lower memory consumption, it may also degrade model performance. The following sections will focus on optimizing GPU memory costs during training in terms of activations and optimizer states.

\begin{figure}[t]
  \centering
  \includegraphics[width=\linewidth]{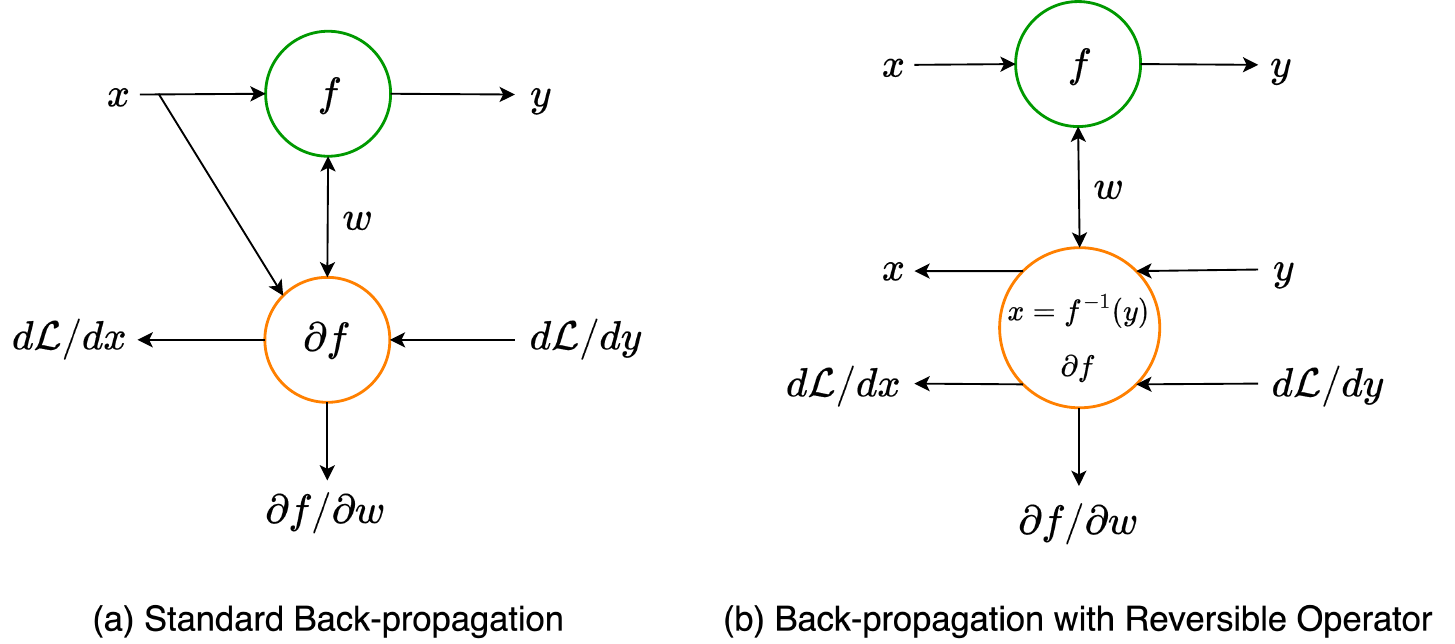}
  \caption{The comparison between reversible and non-reversible operators during the back-propagation.}
  \label{fig:rev_oper}
\end{figure}

\section{Reversible Neural Networks for Deep Speaker Embedding Learning}
\label{sec:revnets}

In this section, we aim to optimize GPU memory usage for activations by designing reversible neural networks. We first introduce a reversible computation principle for speaker verification systems. Then, we apply it to ResNets and DF-ResNets, building two new families of reversible neural network models: RevNets and DF-RevNets. These models can be trained without storing activations during back-propagation, which significantly reduces the GPU memory usage. The details are provided below.

\subsection{Reversible Operator}
\label{ssec:rev_oper}
A reversible operator refers to a transformation that is analytically invertible, meaning the original input can be recovered from the output using its inverse function. Traditional neural networks include irreversible operators, making the entire network non-reversible. For example, the nonlinear ReLU function cannot infer the input from its output. As a result, activations for each computing node must be stored in the back-propagation algorithm. On the contrary, networks built with reversible operators avoid the necessity of caching intermediate activations in GPU memory during the training process. They can recalculate the input from the output during back-propagation and use the recovered inputs to compute gradients for parameter updates. Consequently, reversible neural networks effectively decouple GPU memory costs from network depth, significantly reducing memory requirements when training large-scale networks. Fig. \ref{fig:rev_oper} illustrates the difference between reversible and non-reversible operators in back-propagation.

Consider a simple function $y=f(x)$ as an example. In standard back-propagation, the input $x$ needs to be stored during the forward pass so it can be used directly to compute the gradient, aiming to speed up the training process. However, if the function is reversible, the inverse function $f^{-1}$ can be applied to recover the input $x$ from the output $y$ as $x=f^{-1}(y)$ during the backward pass. This approach allows gradient computation and parameter updates without storing the intermediate activation $x$ during the forward pass, leading to a significant reduction in GPU memory consumption. The trade-off of this method is that it sacrifices time for space, as the extra step of recovering the input increases the overall training time.

In the following, we will design reversible operators and neural networks for speaker verification systems based on ResNets and DF-ResNets to optimize GPU memory usage for activations.

\subsection{RevNets}
\label{ssec:revnets}
In this part, we will first construct reversible neural networks for SV systems based on ResNets, referred to as RevNets. These networks can be divided into two categories based on their degree of reversibility: Type I RevNets and Type II RevNets. The specific content will be presented below.

\subsubsection{Type I RevNets}
Firstly, we provide the details of Type I RevNets. In the original ResNet, He et al.~\cite{ResNet} propose two distinct types of residual blocks: Basic Block and Bottleneck Block. These blocks serve as fundamental components for constructing the network, each comprising a standard convolution operation, batch normalization (BN), and a ReLU activation function. The basic block includes two standard convolution operations, both using a $3 \times 3$ kernel size, while the bottleneck block consists of three convolutions: a $1 \times 1$ convolution, followed by a $3 \times 3$ convolution, and ending with another $1 \times 1$ convolution. The residual block computation can be expressed as:
\begin{align}
  y &= x + \mathcal{F}(x) \label{eq:revnet1}
\end{align}
where $x$ is the activations of the previous layer, $\mathcal{F}$ represents the residual function, which can be either a basic or bottleneck block, and $y$ denotes the activations of the current layer.

The residual function of the basic block, namely $\mathcal{F}_{basic}$, is defined as:
\begin{align}
  \mathcal{F}_{basic} &= Conv(ReLU(BN(Conv(x)))) \label{eq:revnet2}
\end{align}

The residual function of the bottleneck block, namely $\mathcal{F}_{bottleneck}$, is defined as:
\begin{align}
  \mathcal{F}_{bottleneck} &= Conv(Conv(ReLU(BN(Conv(x))))) \label{eq:revnet3}
\end{align}

\begin{table*}[t]
\caption{Type I RevNets: RevNet46 corresponds to ResNet34; RevNet126 and RevNet140 correspond to ResNet101, where RevNet126 uses basic block and RevNet140 utilizes bottleneck block. ``rev\_res'' stands for reversible residual block. ``ds'' is downsampling layer. ``GSP'' denotes global statistical pooling. ``fc'' means fully-connected layer.}
\label{tab:typeI_revnets}
\centering
\begin{adjustbox}{width=.98\textwidth,center}
\begin{tabular}{c|c|c|c}
  \toprule
  \textbf{Stage} & \textbf{RevNet46} & \textbf{RevNet126}  & \textbf{RevNet140}     \\
  \hline
  conv1 & $3\times3, 48, \text{stride 1}$ & $3\times3, 48, \text{stride 1}$  & $3\times3, 48, \text{stride 1}$ \\
  \hline
  res2 & $\begin{bmatrix} 3\times3, 48 \\ 3\times3, 48 \end{bmatrix} \times 1$ & $\begin{bmatrix} 3\times3, 48 \\ 3\times3, 48 \end{bmatrix} \times 1$ & $\begin{bmatrix} 1\times1, 48  \\ 3\times3, 48 \\ 1\times1, 192 \end{bmatrix} \times 1$ \\
  \hline
  rev\_res3 & $\mathcal{F}:$ $\begin{bmatrix} 3\times3, 24 \\ 3\times3, 24 \end{bmatrix} \times 1$ $\mathcal{G}:$ $\begin{bmatrix} 3\times3, 24 \\ 3\times3, 24 \end{bmatrix} \times 1$  & $\mathcal{F}:$ $\begin{bmatrix} 3\times3, 24 \\ 3\times3, 24 \end{bmatrix} \times 2$ $\mathcal{G}:$ $\begin{bmatrix} 3\times3, 24 \\ 3\times3, 24 \end{bmatrix} \times 2$  & $\mathcal{F}:$ $\begin{bmatrix} 1\times1, 24  \\ 3\times3, 24 \\ 1\times1, 96 \end{bmatrix} \times 2$ $\mathcal{G}:$ $\begin{bmatrix} 1\times1, 24  \\ 3\times3, 24 \\ 1\times1, 96 \end{bmatrix} \times 2$ \\
  \hline
  ds4 & $\begin{bmatrix} 3\times3, 96 \\ 3\times3, 96 \end{bmatrix} \times 1$ & $\begin{bmatrix} 3\times3, 96 \\ 3\times3, 96 \end{bmatrix} \times 1$ & $\begin{bmatrix} 1\times1, 96  \\ 3\times3, 96 \\ 1\times1, 384 \end{bmatrix} \times 1$ \\
  \hline
  rev\_res5 & $\mathcal{F}:$ $\begin{bmatrix} 3\times3, 48 \\ 3\times3, 48 \end{bmatrix} \times 2$ $\mathcal{G}:$ $\begin{bmatrix} 3\times3, 48 \\ 3\times3, 48 \end{bmatrix} \times 2$  & $\mathcal{F}:$ $\begin{bmatrix} 3\times3, 48 \\ 3\times3, 48 \end{bmatrix} \times 3$ $\mathcal{G}:$ $\begin{bmatrix} 3\times3, 48 \\ 3\times3, 48 \end{bmatrix} \times 3$ & $\mathcal{F}:$ $\begin{bmatrix} 1\times1, 48  \\ 3\times3, 48 \\ 1\times1, 192 \end{bmatrix} \times 3$ $\mathcal{G}:$ $\begin{bmatrix} 1\times1, 48  \\ 3\times3, 48 \\ 1\times1, 192 \end{bmatrix} \times 3$ \\
  \hline
  ds6 & $\begin{bmatrix} 3\times3, 192 \\ 3\times3, 192 \end{bmatrix} \times 1$ & $\begin{bmatrix} 3\times3, 192 \\ 3\times3, 192 \end{bmatrix} \times 1$ & $\begin{bmatrix} 1\times1, 192  \\ 3\times3, 192 \\ 1\times1, 768 \end{bmatrix} \times 1$ \\
  \hline
  rev\_res7 & $\mathcal{F}:$ $\begin{bmatrix} 3\times3, 96 \\ 3\times3, 96 \end{bmatrix} \times 4$ $\mathcal{G}:$ $\begin{bmatrix} 3\times3, 96 \\ 3\times3, 96 \end{bmatrix} \times 4$  & $\mathcal{F}:$ $\begin{bmatrix} 3\times3, 96 \\ 3\times3, 96 \end{bmatrix} \times 22$ $\mathcal{G}:$ $\begin{bmatrix} 3\times3, 96 \\ 3\times3, 96 \end{bmatrix} \times 22$ & $\mathcal{F}:$ $\begin{bmatrix} 1\times1, 96  \\ 3\times3, 96 \\ 1\times1, 384 \end{bmatrix} \times 14$ $\mathcal{G}:$ $\begin{bmatrix} 1\times1, 96  \\ 3\times3, 96 \\ 1\times1, 384 \end{bmatrix} \times 14$ \\
  \hline
  ds8 & $\begin{bmatrix} 3\times3, 300 \\ 3\times3, 300 \end{bmatrix} \times 1$ & $\begin{bmatrix} 3\times3, 384 \\ 3\times3, 384 \end{bmatrix} \times 1$ & $\begin{bmatrix} 1\times1, 300  \\ 3\times3, 300 \\ 1\times1, 1200 \end{bmatrix} \times 1$ \\
  \hline
  rev\_res9 & $\mathcal{F}:$ $\begin{bmatrix} 3\times3, 150 \\ 3\times3, 150 \end{bmatrix} \times 2$ $\mathcal{G}:$ $\begin{bmatrix} 3\times3, 150 \\ 3\times3, 150 \end{bmatrix} \times 2$  & $\mathcal{F}:$ $\begin{bmatrix} 3\times3, 192 \\ 3\times3, 192 \end{bmatrix} \times 2$ $\mathcal{G}:$ $\begin{bmatrix} 3\times3, 192 \\ 3\times3, 192 \end{bmatrix} \times 2$ & $\mathcal{F}:$ $\begin{bmatrix} 1\times1, 150  \\ 3\times3, 150 \\ 1\times1, 600 \end{bmatrix} \times 2$ $\mathcal{G}:$ $\begin{bmatrix} 1\times1, 150  \\ 3\times3, 150 \\ 1\times1, 600 \end{bmatrix} \times 2$ \\
  \hline
  pooling  & GSP  & GSP & GSP \\
  \hline
  fc  & $\text{(6000, 256)}$  & $\text{(7680, 256)}$ & $\text{(24000, 256)}$ \\
  \hline
  \# params & $6.7 \times 10^6$  & $15.0 \times 10^6$  & $15.8 \times 10^6$  \\
  \bottomrule
\end{tabular}
\end{adjustbox}
\end{table*}

\begin{algorithm}[t]
  \caption{Back-propagation for Reversible Blocks}
  \label{alg:backprob}
  \begin{algorithmic}[1]
    \Require
      $(y_1, y_2)$;
      $(d\mathcal{L} / dy_1, d\mathcal{L} / dy_2)$
    \Ensure
      $(x_1, x_2)$;
      $(d\mathcal{L} / dx_1, d\mathcal{L} / dx_2)$;
      $(d\mathcal{L} / d w_{\mathcal{F}}, d\mathcal{L} / d w_{\mathcal{G}})$
    \State $z_1 \gets y_1$;
    \State $x_2 \gets y_2 - \mathcal{G}(z_1)$;
    \State $x_1 \gets z_1 - \mathcal{F}(x_2)$;
    \State $\frac{d\mathcal{L}}{dz_1} \gets \frac{d\mathcal{L}}{dy_1} + \Bigl(\frac{\partial \mathcal{G}}{\partial z_1}\Bigr)^T \frac{d\mathcal{L}}{dy_2}$;
    \State $\frac{d\mathcal{L}}{dx_2} \gets \frac{d\mathcal{L}}{dy_2} + \Bigl(\frac{\partial \mathcal{F}}{\partial x_2}\Bigr)^T \frac{d\mathcal{L}}{dz_1}$;
    \State $\frac{d\mathcal{L}}{dx_1} \gets \frac{d\mathcal{L}}{dz_1}$;
    \State $\frac{d\mathcal{L}}{dw_{\mathcal{F}}} \gets \Bigl(\frac{\partial \mathcal{F}}{\partial w_{\mathcal{F}}}\Bigr)^T \frac{d\mathcal{L}}{dz_1}$;
    \State $\frac{d\mathcal{L}}{dw_{\mathcal{G}}} \gets \Bigl(\frac{\partial \mathcal{G}}{\partial w_{\mathcal{G}}}\Bigr)^T \frac{d\mathcal{L}}{dy_2}$;
    
  \end{algorithmic}
\end{algorithm}

The primary challenge in constructing a reversible version of ResNet lies in how to convert the non-reversible residual block into a reversible function. Inspired by ~\cite{revnet}, we first split the input activation $x$ evenly into $x_1$ and $x_2$ along the channel dimension. Using the additive coupling mechanism, the original residual block is transformed into a reversible function. The calculation process is as follows:
\begin{align}
  y_1 &= x_1 + \mathcal{F}(x_2) \nonumber \\
  y_2 &= x_2 + \mathcal{G}(y_1) \label{eq:revnet4}
\end{align}
where $x_1$ and $x_2$ are the split activations, $y_1$ and $y_2$ are the corresponding outputs. $\mathcal{F}$ and $\mathcal{G}$ represent the residual functions mentioned above, which can either be a basic or a bottleneck block.

During the back-propagation, the input activations $x_1$ and $x_2$ can be recovered from the outputs $y_1$ and $y_2$ using the reversible function. The formula is as follows:
\begin{align}
  z_1 &= y_1 \nonumber \\
  x_2 &= y_2 - \mathcal{G}(z_1) \nonumber \\
  x_1 &= z_1 - \mathcal{F}(x_2) \label{eq:revnet5}
\end{align}

Reversibility is maintained only when the convolution operations in the residual blocks $\mathcal{F}$ and $\mathcal{G}$ have a stride of 1. If the stride exceeds 1, the layer's computation loses information, making it irreversible. In the original ResNet, spatial downsampling is performed at the beginning of each computational phase, typically using convolution operations with a stride of 2. As a result, the described method can not make these downsampled layers reversible. Therefore, we preserve these irreversible downsampled layers, creating Type I RevNets with partially non-reversible layers.

For reversible blocks, back-propagation can be performed without the need to store intermediate activations. Given activations $y_1$ and $y_2$ along with their associated total derivatives, the input activations $x_1$ and $x_2$ can be recovered first. Then, the total derivatives concerning $x_1$, $x_2$, and the parameters involved in $\mathcal{F}$ and $\mathcal{G}$ are computed through Alg. \ref{alg:backprob}. For non-reversible layers, activations must be stored explicitly to calculate gradients. However, since there are only a few such layers, the overall memory usage is significantly reduced and remains constant regardless of network depth. As a result, the storage cost of activations is independent of depth. Notably, Alg. \ref{alg:backprob} can be applied to any residual function for $\mathcal{F}$ and $\mathcal{G}$. Although the algorithm requires some manual steps during back-propagation, different residual functions can be directly substituted without modifying the implementation.

Based on the above reversible blocks, we construct Type I RevNets for ResNet34, ResNet101 and ResNet152. To ensure a fair comparison, we aim to maintain a similar number of parameters between the original models and their reversible counterparts. Table \ref{tab:typeI_revnets} displays Type I RevNets corresponding to ResNet34 and ResNet101. For ResNet152, the reversible versions follow the same architecture as RevNet126 and RevNet140, but with a different number of blocks. The specific settings are listed below.
\begin{itemize}
  \item RevNet178: $C$=[48, 96, 192, 384], $B$=[3, 8, 32, 3]
  \item RevNet230: $C$=[48, 96, 192, 300], $B$=[3, 8, 26, 3]
\end{itemize}

\begin{table*}[t]
\caption{Type II RevNets: RevNet57 corresponds to ResNet34; RevNet137 and RevNet155 correspond to ResNet101, where RevNet137 uses basic block and RevNet155 adopts bottleneck block. ``rev\_ds'' stands for reversible dowmsampling layer.}
\label{tab:typeII_revnets}
\centering
\begin{adjustbox}{width=.98\textwidth,center}
\begin{tabular}{c|c|c|c}
  \toprule
  \textbf{Stage} & \textbf{RevNet57} & \textbf{RevNet137}  & \textbf{RevNet155}     \\
  \hline
  conv1 & $3\times3, 48, \text{stride 1}$ & $3\times3, 48, \text{stride 1}$  & $3\times3, 48, \text{stride 1}$ \\
  \hline
  rev\_res2 & $\mathcal{F}:$ $\begin{bmatrix} 3\times3, 24 \\ 3\times3, 24 \end{bmatrix} \times 2$ $\mathcal{G}:$ $\begin{bmatrix} 3\times3, 24 \\ 3\times3, 24 \end{bmatrix} \times 2$  & $\mathcal{F}:$ $\begin{bmatrix} 3\times3, 24 \\ 3\times3, 24 \end{bmatrix} \times 3$ $\mathcal{G}:$ $\begin{bmatrix} 3\times3, 24 \\ 3\times3, 24 \end{bmatrix} \times 3$  & $\mathcal{F}:$ $\begin{bmatrix} 1\times1, 24  \\ 3\times3, 24 \\ 1\times1, 96 \end{bmatrix} \times 3$ $\mathcal{G}:$ $\begin{bmatrix} 1\times1, 24  \\ 3\times3, 24 \\ 1\times1, 96 \end{bmatrix} \times 3$ \\
  \hline
  conv3 & $3\times3, 24, \text{stride 1}$ & $3\times3, 24, \text{stride 1}$ & $3\times3, 24, \text{stride 1}$  \\
  \hline
  rev\_ds4 & $ \text{tensor reshape (r=2)}, 96$ & $ \text{tensor reshape (r=2)}, 96$ & $ \text{tensor reshape (r=2)}, 96$  \\
  \hline
  rev\_res5 & $\mathcal{F}:$ $\begin{bmatrix} 3\times3, 48 \\ 3\times3, 48 \end{bmatrix} \times 3$ $\mathcal{G}:$ $\begin{bmatrix} 3\times3, 48 \\ 3\times3, 48 \end{bmatrix} \times 3$  & $\mathcal{F}:$ $\begin{bmatrix} 3\times3, 48 \\ 3\times3, 48 \end{bmatrix} \times 4$ $\mathcal{G}:$ $\begin{bmatrix} 3\times3, 48 \\ 3\times3, 48 \end{bmatrix} \times 4$ & $\mathcal{F}:$ $\begin{bmatrix} 1\times1, 48  \\ 3\times3, 48 \\ 1\times1, 192 \end{bmatrix} \times 4$ $\mathcal{G}:$ $\begin{bmatrix} 1\times1, 48  \\ 3\times3, 48 \\ 1\times1, 192 \end{bmatrix} \times 4$ \\
  \hline
  conv6 & $3\times3, 48, \text{stride 1}$ & $3\times3, 48, \text{stride 1}$ & $3\times3, 48, \text{stride 1}$ \\
  \hline
  rev\_ds7 & $ \text{tensor reshape (r=2)}, 192$ & $ \text{tensor reshape (r=2)}, 192$ & $ \text{tensor reshape (r=2)}, 192$  \\
  \hline
  rev\_res8 & $\mathcal{F}:$ $\begin{bmatrix} 3\times3, 96 \\ 3\times3, 96 \end{bmatrix} \times 5$ $\mathcal{G}:$ $\begin{bmatrix} 3\times3, 96 \\ 3\times3, 96 \end{bmatrix} \times 5$  & $\mathcal{F}:$ $\begin{bmatrix} 3\times3, 96 \\ 3\times3, 96 \end{bmatrix} \times 23$ $\mathcal{G}:$ $\begin{bmatrix} 3\times3, 96 \\ 3\times3, 96 \end{bmatrix} \times 23$ & $\mathcal{F}:$ $\begin{bmatrix} 1\times1, 96  \\ 3\times3, 96 \\ 1\times1, 384 \end{bmatrix} \times 15$ $\mathcal{G}:$ $\begin{bmatrix} 1\times1, 96  \\ 3\times3, 96 \\ 1\times1, 384 \end{bmatrix} \times 15$ \\
  \hline
  conv9 & $3\times3, 75, \text{stride 1}$ & $3\times3, 96, \text{stride 1}$ & $3\times3, 75, \text{stride 1}$ \\
  \hline
  rev\_ds10 & $ \text{tensor reshape (r=2)}, 300$ & $ \text{tensor reshape (r=2)}, 384$ & $ \text{tensor reshape (r=2)}, 300$  \\
  \hline
  rev\_res11 & $\mathcal{F}:$ $\begin{bmatrix} 3\times3, 150 \\ 3\times3, 150 \end{bmatrix} \times 3$ $\mathcal{G}:$ $\begin{bmatrix} 3\times3, 150 \\ 3\times3, 150 \end{bmatrix} \times 3$  & $\mathcal{F}:$ $\begin{bmatrix} 3\times3, 192 \\ 3\times3, 192 \end{bmatrix} \times 3$ $\mathcal{G}:$ $\begin{bmatrix} 3\times3, 192 \\ 3\times3, 192 \end{bmatrix} \times 3$ & $\mathcal{F}:$ $\begin{bmatrix} 1\times1, 150  \\ 3\times3, 150 \\ 1\times1, 600 \end{bmatrix} \times 3$ $\mathcal{G}:$ $\begin{bmatrix} 1\times1, 150  \\ 3\times3, 150 \\ 1\times1, 600 \end{bmatrix} \times 3$ \\
  \hline
  pooling  & GSP  & GSP & GSP \\
  \hline
  fc  & $\text{(6000, 256)}$  & $\text{(7680, 256)}$ & $\text{(24000, 256)}$ \\
  \hline
  \# params & $6.1 \times 10^6$  & $14.2 \times 10^6$  & $15.6 \times 10^6$  \\
  \bottomrule
\end{tabular}
\end{adjustbox}
\end{table*}

\subsubsection{Type II RevNets}
As mentioned earlier, traditional spatial downsampling methods, like convolutions with a stride greater than 1 or max pooling, are inherently irreversible because they alter the spatial dimensions of input activations, leading to information loss. Consequently, in Type I RevNets, the downsampled activations must still be cached in the GPU during the forward pass. To further increase the level of reversibility, we introduce Type II RevNets by making the downsampling operations invertible.

Inspired by a pixel compression operation for image data, this method first divides the image into sub-rectangular blocks of shape $C \times 2 \times 2$ per channel and then reshapes these blocks into a $4C \times 1 \times 1$ format. This operation compresses a tensor of shape $C \times H \times W$ into $4C \times H/2 \times W/2$, effectively reducing the spatial dimensions while increasing the number of channels in a reversible manner. We extend this concept to audio data by rearranging the original speech features of shape $C \times F \times T$ (where $C$, $F$, and $T$ represent the channel, frequency, and time dimensions) using a ratio $r$.
\begin{align}
  C \times F \times T \rightarrow r^2C \times F/r \times T/r \label{eq:revnet6}
\end{align}

This operation achieves downsampling while maintaining the total number of elements in the feature tensor and ensuring reversibility. By replacing the non-reversible downsampling layer in Type I RevNets with this reversible compression operation, we obtain Type II RevNets. Table \ref{tab:typeII_revnets} presents the corresponding Type II RevNets for ResNet34 and ResNet101. For ResNet152, the reversible variants share the same structure as RevNet137 and RevNet155, differing only in the number of blocks. The detailed configurations are provided below.
\begin{itemize}
  \item RevNet197: $C$=[48, 96, 192, 384], $B$=[3, 8, 34, 3]
  \item RevNet245: $C$=[48, 96, 192, 300], $B$=[3, 8, 26, 3]
\end{itemize}

\subsection{DF-RevNets}
\label{ssec:df_revnets}
In this part, we extend the above concept to DF-ResNets, with their reversible variants named as DF-RevNets. Similar to the previous RevNets, these networks are categorized into two types: Type I DF-RevNets and Type II DF-RevNets. The following will provide detailed explanations.

\subsubsection{Type I DF-RevNets}
Unlike ResNets, DF-ResNets only contain one type of residual block, known as the DF-Bottleneck. This block is composed of three convolutional layers: a $1 \times 1$ convolution, followed by a $3 \times 3$ depth-wise convolution, and another $1 \times 1$ convolution. The residual function $\mathcal{F}_{df-bottleneck}$ is defined as follows:
\begin{align}
  \mathcal{F}_{df-bottleneck} &= Conv(DConv(ReLU(BN(Conv(x))))) \label{eq:revnet7}
\end{align}

As previously mentioned, the reversible operation is independent of the specific form of the residual function, allowing the DF-Bottleneck to be converted into a reversible block using Eq. \ref{eq:revnet4} and Eq. \ref{eq:revnet5}. Meanwhile, we retain the irreversible downsampling layer and construct Type I DF-RevNets for DF-ResNet56, DF-ResNet110, DF-ResNet179, and DF-ResNet233. Table \ref{tab:dfresnet56} shows the Type I RevNets corresponding to DF-ResNet56. The reversible versions of DF-ResNet110, DF-ResNet179, and DF-ResNet233 follow the same architecture as DF-RevNet66, only with a different number of blocks. Detailed configurations are provided below.
\begin{itemize}
  \item DF-RevNet126: $C$=[48, 96, 192, 384], $B$=[3, 3, 15, 3]
  \item DF-RevNet258: $C$=[48, 96, 192, 384], $B$=[3, 8, 32, 3]
  \item DF-RevNet354: $C$=[48, 96, 192, 384], $B$=[3, 8, 48, 3]
\end{itemize}

\subsubsection{Type II DF-RevNets}
Similar to ResNet, Type II DF-RevNets can be derived by substituting the irreversible downsampling layer in Type I DF-RevNets with the reversible compression operation described in Eq. \ref{eq:revnet6}, we can derive Type II DF-RevNets. Table \ref{tab:dfresnet56} lists the Type II reversible version of DF-ResNet56. For DF-ResNet110, DF-ResNet179, and DF-ResNet233, their reversible variants have the same structure as DF-RevNet89, with only different block numbers. The configurations are as follows.
\begin{itemize}
  \item DF-RevNet149: $C$=[48, 96, 192, 384], $B$=[3, 3, 15, 3]
  \item DF-RevNet281: $C$=[48, 96, 192, 384], $B$=[3, 8, 32, 3]
  \item DF-RevNet377: $C$=[48, 96, 192, 384], $B$=[3, 8, 48, 3]
\end{itemize}

\begin{table}[t]
\caption{The reversible variants of DF-ResNet56: DF-RevNet66 is Type I DF-RevNets and DF-RevNet89 is Type II DF-RevNets.}
\label{tab:dfresnet56}
\centering
\begin{adjustbox}{width=.48\textwidth,center}
\begin{tabular}{c|c|c|c}
  \toprule
  \textbf{Stage} & \textbf{DF-RevNet66 (Type I)}  & \textbf{DF-RevNet89 (Type II)} & \textbf{Stage}    \\
  \hline
  conv1 \& 2 & $3\times3, 48, \text{stride 1}$  & $3\times3, 48, \text{stride 1}$ & conv1 \\
  \hline
  rev\_res3 & $\mathcal{F}:$ $\begin{bmatrix} 1\times1, 96  \\ \text{d}3\times3, 96 \\ 1\times1, 24 \end{bmatrix} \times 2$ $\mathcal{G}:$ $\begin{bmatrix} 1\times1, 96  \\ \text{d}3\times3, 96 \\ 1\times1, 24 \end{bmatrix} \times 2$ & $\mathcal{F}:$ $\begin{bmatrix} 1\times1, 96  \\ \text{d}3\times3, 96 \\ 1\times1, 24 \end{bmatrix} \times 3$ $\mathcal{G}:$ $\begin{bmatrix} 1\times1, 96  \\ \text{d}3\times3, 96 \\ 1\times1, 24 \end{bmatrix} \times 3$ & rev\_res2 \\
  \hline
  ds4 & $3\times3, 96, \text{stride 2}$  & $3\times3, 24, \text{stride 1}$ & conv3\\
  \hline
  -- & -- & tensor reshape (r=2), 96 & rev\_ds4 \\
  \hline
  rev\_res5 & $\mathcal{F}:$ $\begin{bmatrix} 1\times1, 192  \\ \text{d}3\times3, 192 \\ 1\times1, 48 \end{bmatrix} \times 2$ $\mathcal{G}:$ $\begin{bmatrix} 1\times1, 192  \\ \text{d}3\times3, 192 \\ 1\times1, 48 \end{bmatrix} \times 2$ & $\mathcal{F}:$ $\begin{bmatrix} 1\times1, 192  \\ \text{d}3\times3, 192 \\ 1\times1, 48 \end{bmatrix} \times 3$ $\mathcal{G}:$ $\begin{bmatrix} 1\times1, 192  \\ \text{d}3\times3, 192 \\ 1\times1, 48 \end{bmatrix} \times 3$ & rev\_res5 \\
  \hline
  ds6 & $3\times3, 192, \text{stride 2}$  & $3\times3, 48, \text{stride 1}$ & conv6 \\
  \hline
  -- & -- & tensor reshape (r=2), 192 & rev\_ds7 \\
  \hline
  rev\_res7 & $\mathcal{F}:$ $\begin{bmatrix} 1\times1, 384  \\ \text{d}3\times3, 384 \\ 1\times1, 96 \end{bmatrix} \times 4$ $\mathcal{G}:$ $\begin{bmatrix} 1\times1, 384  \\ \text{d}3\times3, 384 \\ 1\times1, 96 \end{bmatrix} \times 4$ & $\mathcal{F}:$ $\begin{bmatrix} 1\times1, 384  \\ \text{d}3\times3, 384 \\ 1\times1, 96 \end{bmatrix} \times 5$ $\mathcal{G}:$ $\begin{bmatrix} 1\times1, 384  \\ \text{d}3\times3, 384 \\ 1\times1, 96 \end{bmatrix} \times 5$ & rev\_res8 \\
  \hline
  ds8 & $3\times3, 384, \text{stride 2}$  & $3\times3, 96, \text{stride 1}$ & conv9\\
  \hline
  -- & -- & tensor reshape (r=2), 384 & rev\_ds10 \\
  \hline
  rev\_res9 & $\mathcal{F}:$ $\begin{bmatrix} 1\times1, 768  \\ \text{d}3\times3, 768 \\ 1\times1, 192 \end{bmatrix} \times 2$ $\mathcal{G}:$ $\begin{bmatrix} 1\times1, 768  \\ \text{d}3\times3, 768 \\ 1\times1, 192 \end{bmatrix} \times 2$ & $\mathcal{F}:$ $\begin{bmatrix} 1\times1, 768  \\ \text{d}3\times3, 768 \\ 1\times1, 192 \end{bmatrix} \times 3$ $\mathcal{G}:$ $\begin{bmatrix} 1\times1, 768  \\ \text{d}3\times3, 768 \\ 1\times1, 192 \end{bmatrix} \times 3$ & rev\_res11 \\
  \hline
  pooling  & GSP & GSP & pooling \\
  \hline
  fc  & $\text{(7680, 256)}$ & $\text{(7680, 256)}$ & fc \\
  \hline
  \# params  & $4.8 \times 10^6$  & $4.5 \times 10^6$ & \# params \\
  \bottomrule
\end{tabular}
\end{adjustbox}
\end{table}

\section{Optimizer State Quantization for Deep Speaker Embedding Learning}
\label{sec:optimizer_quant}
In this section, we focus on reducing GPU memory consumption for optimizer states by introducing a dynamic quantization method. We begin by analyzing the state storage requirements of commonly used optimizers like SGD and Adam. Next, we propose an 8-bit dynamic quantization algorithm for optimizer states, which significantly reduces GPU memory usage without impacting the precision of weight updates. The details are explained below.

\subsection{Optimizer State Analysis}
As discussed in Section \ref{ssec:back_prob}, during back-propagation, the optimizer uses the gradient $\mathbf{g}_t = d\mathcal{L} / d\mathbf{w}$ at time step $t$ to update the neural network parameters $\mathbf{w}$. Additionally, stateful optimizers maintain gradient statistics for each weight over time to enhance training stability and accelerate convergence. For speaker verification systems, the most popular stateful optimizers are SGD with momentum and Adam, along with its variant AdamW. In this part, we first examine the update rules and state storage of these optimizers. For SGD, the update process is as follows:
\begin{align}
  \text{SGD}(\mathbf{g}_t,\mathbf{w}_{t-1}, \mathbf{m}_{t-1}) &=  
  \begin{cases}
    \mathbf{m}_0 = \mathbf{g}_0 \\
    \mathbf{m}_t = \beta\mathbf{m}_{t-1} +\mathbf{g}_t \\
    \mathbf{w}_t = \mathbf{w}_{t-1} - \alpha\cdot\mathbf{m}_t
  \end{cases} \label{eq:sgd}
\end{align}
where $\mathbf{w}_{t-1}$ denotes the weight at time step $t-1$, $\mathbf{m}_{t-1}$ represents the optimizer state at time step $t-1$, $\mathbf{g}_t$ is the gradient at time step $t$, $\alpha$ is the learning rate, and $\beta$ is the momentum coefficient. Hence, the SGD optimizer with momentum requires storing both the gradient $\mathbf{g}_t$ and the state $\mathbf{m}_t$, which are typically represented as 32-bit floating-point numbers by default.

For Adam and its variant AdamW, the calculation process is as follows:
\begin{small}
\begin{align}
  \text{Adam}(\mathbf{g}_t,\mathbf{w}_{t-1}, \mathbf{m}_{t-1}, \mathbf{r}_{t-1}) &=  
  \begin{cases}
    \mathbf{r}_0 = \mathbf{m}_0 = \mathbf{0} \\
    \mathbf{m}_t = \beta_1\mathbf{m}_{t-1} + (1-\beta_1)\mathbf{g}_t \\
    \mathbf{r}_t = \beta_2\mathbf{r}_{t-1} + (1-\beta_2)\mathbf{g}_t^2 \\
    \mathbf{w}_t = \mathbf{w}_{t-1} - \alpha\cdot\frac{\mathbf{m}_t}{\sqrt{\mathbf{r}_t}+\epsilon}
  \end{cases}  \label{eq:adam}
\end{align}
\end{small}
where $\mathbf{m}_{t-1}$ represents the first-order optimizer state at time step $t-1$, and $\mathbf{r}_{t-1}$ is the second-order optimizer state. The coefficients $\beta_1$ and $\beta_2$ correspond to the statistics of state 1 and state 2, respectively. As a result, the Adam optimizer needs to store gradient $\mathbf{g}_t$ along with two gradient statistics, $\mathbf{m}_t$  $\mathbf{r}_t$.

\subsection{8-bit Dynamic Quantization}
To reduce the GPU memory usage of the optimizer state, we introduce a dynamic quantization algorithm, which can convert the original 32-bit floating-point numbers into 8-bit integers for storage and dequantize them back to 32-bit values during parameter updates. This approach effectively decreases GPU memory consumption for the optimizer state without affecting model performance. Fig. \ref{fig:8bit_quant} depicts the process of 8-bit dynamic quantization and dequantization. The algorithm is detailed as follows:

\begin{figure}[t]
  \centering
  \includegraphics[width=\linewidth]{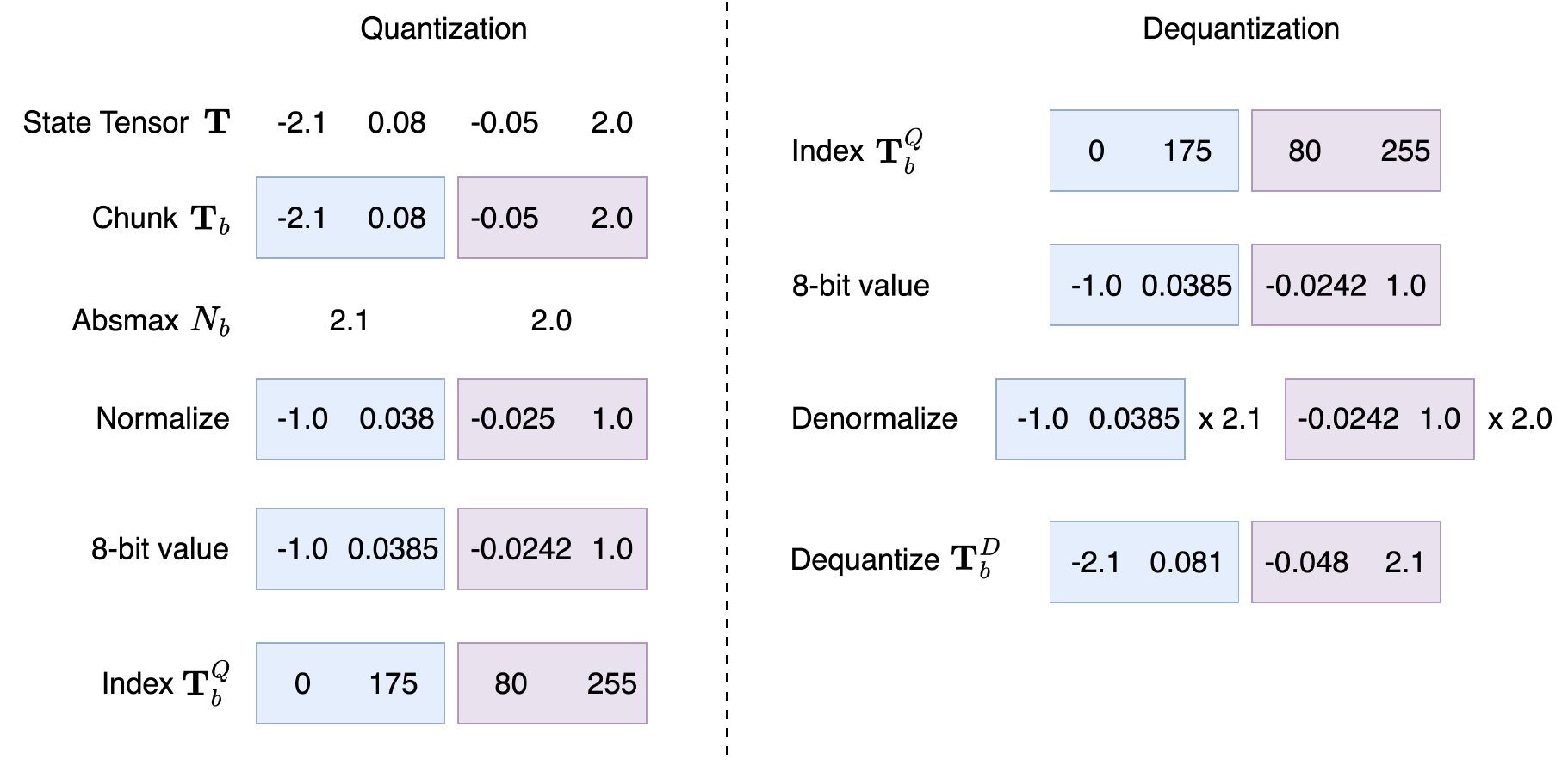}
  \caption{illustration of 8-bit dynamic quantization and dequantization process for optimizer state.}
  \label{fig:8bit_quant}
\end{figure}

\subsubsection{Tensor chunking}
To mitigate the impact of outliers on quantization precision, we first partition the optimizer state tensor into smaller blocks. Specifically, the tensor is divided into blocks of size $B$, with quantization applied independently to each block. In this process, the state tensor $\mathbf{T}$ is flattened into a one-dimensional sequence and then segmented into $B$ blocks. For a state tensor $\mathbf{T}$ with $n$ elements, this results in $n/B$ blocks. In the experiment, we set $B$ to 2048.

\subsubsection{Dynamic quantization}
Next, we perform 8-bit dynamic quantization independently within each block. The quantization process maps $n$-bit integers to a real-valued domain $D$, expressed as ${\mathbf{Q}^{\text{map}}: [0, 2^n-1] \mapsto D}$. We denote this mapping as ${\mathbf{Q}^{\text{map}}(i) = \mathbf{Q}^{\text{map}}_i = q_i}$, where integer $i$ is mapped to the floating-point value $q$. To quantize 32-bit floating-point data to an 8-bit format, the process involves the following three steps:

\textit{Normalization: }For each block $\mathbf{T}_b$, we first compute the maximum absolute value $N_b$ and then normalize it.
\begin{align}
  N_b &= \max(|\mathbf{T}_b|) \nonumber \\
  & \mathbf{T}_b \gets \frac{\mathbf{T}_b}{N_b} \label{eq:quant1}
\end{align}
where $b$ is the block index and $0 \leq b \leq n/B$.

\begin{figure}[t]
  \centering
  \includegraphics[width=\linewidth]{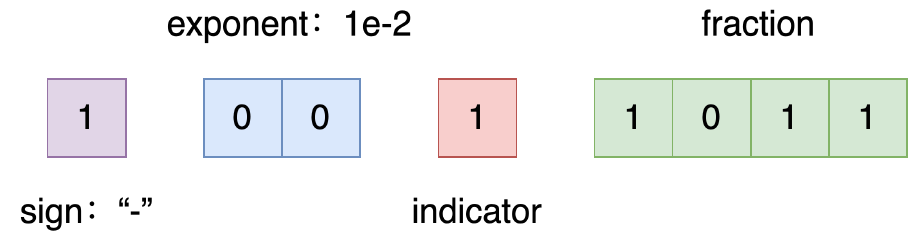}
  \caption{8-bit dynamic tree quantization data type.}
  \label{fig:dynamic_tree}
\end{figure}

\textit{Nearest value searching: }For each 32-bit floating-point number in the block $\mathbf{T}_b$, we identify the nearest value $q_i$ using binary search within the real-number domain $D$ of the 8-bit data type. This process employs the dynamic tree quantization method~\cite{dynamic-tree}, which implements an 8-bit floating-point format capable of preserving quantization accuracy for both small and large magnitude values. Unlike formats with fixed exponent and fraction components, dynamic tree quantization adjusts these components adaptively for each number. As illustrated in Fig. \ref{fig:dynamic_tree}, this method is composed of four main elements: (1) the first bit indicates the sign of the number; (2) the subsequent series of zero bits denotes the exponent's power; (3) an indicator bit set to 1 signifies the start of the fractional part; (4) the remaining bits represent the fractional component. By adjusting the indicator bits, this format supports exponents as small as $10^{-7}$ or fractional precision up to $1/63$. Dynamic tree quantization can represent values in the range $[-1.0, 1.0]$, offering lower absolute and relative quantization error for non-uniformly distributed data.

\textit{Indexing: }Finally, we store the integer index $i$ corresponding to $q_i$ as the quantization result of block $\mathbf{T}_b$, denoted as $\mathbf{T}_b^Q$. The whole process can be formulated as follows:
\begin{align}
  \mathbf{T}_{bi}^Q &= \argmin_{j=0}^{2^n-1} |\mathbf{{Q}}^{\text{map}}_j - \frac{\mathbf{T}_{bi}}{N_b}|{\bigg\rvert}_{0<i<B} \label{eq:quant2}
\end{align}

\subsubsection{Dequantization}
After obtaining the quantized integer $\mathbf{T}_b^Q$, the dequantization process involves the following steps, as illustrated in Fig. \ref{fig:8bit_quant}: first, retrieve the stored index $i$, then map it back to the 8-bit floating-point format, and finally denormalize to obtain the dequantized tensor $\mathbf{T}_b^D$. The detailed calculation process is as follows:
\begin{align}
  \mathbf{T}_{bi}^D &= {\mathbf{Q}^{\text{map}}}(\mathbf{T}_{bi}^Q) \cdot N_b {\bigg\rvert}_{0<i<B} \label{eq:quant3}
\end{align}

We apply the 8-bit dynamic quantization algorithm to both traditional SGD and Adam, yielding the corresponding 8-bit versions of the optimizers: SGD (8-bit) and Adam (8-bit).

\section{Experimental Setups}
\subsection{Datasets and Data Augmentation}
To examine the performance of our proposed approaches, we employ the Voxceleb1 and Voxceleb2 benchmark datasets~\cite{voxceleb1, voxceleb2} for speaker verification tasks. These datasets consist of audio segments extracted from interviews with over 6,000 celebrities on YouTube. For training, all models utilize the Voxceleb2 development set, comprising roughly 2,200 hours of speech data, which includes around 1 million utterances from 5,994 speakers. Testing is performed using the official Voxceleb evaluation trials: Vox1-O, Vox1-E, and Vox1-H. Additionally, to generate more diverse training samples, we hypothesize that changes in speech speed can result in new speakers~\cite{speed_perturb}. By adjusting the speed to 0.9x or 1.1x, we effectively triple the number of speakers in the dataset. Furthermore, we apply online data augmentation~\cite{online_data_aug} by randomly adding background noise from the MUSAN~\cite{musan} and reverberation from the RIR~\cite{rir} to the training speech, with a probability of 0.6.

\begin{table*}[t]
  \caption{Architectural details and EER (\%) results of ResNets and our proposed RevNets on the Voxceleb1 dataset. All memory usage and maximum batch size are measured on a single 11GB 2080Ti GPU with 2-seconds utterances.}
  \label{tab:revnets}
  \centering
  \setlength{\doublerulesep}{4.5pt}
  \begin{adjustbox}{width=.98\textwidth,center}
  \begin{tabular}{|l|c|c|c|c|c|c|c|c|c|}
    \hline
    \multirow{2}{*}{\textbf{System}} & \multirow{2}{*}{\textbf{Reversible}} & \multirow{2}{*}{\textbf{Block}} & \multirow{2}{*}{\textbf{\# Params}} & \multirow{2}{*}{\textbf{Memory (GB/utter)}} & \multirow{2}{*}{\textbf{Compression Ratio}} & \multirow{2}{*}{\textbf{Maximum Batch Size}} & \multirow{2}{*}{\textbf{Vox1-O}} & \multirow{2}{*}{\textbf{Vox1-E}} & \multirow{2}{*}{\textbf{Vox1-H}} \\ 
    & & & & & & & & &     \\
    \hline
    \hline
    ResNet34 & -- & \multirow{5}{*}{Basic}  & 6.6M & 0.06 & -- & 154 & 0.96 & 1.01 & 1.86 \\
    \cline{1-2}
    \cline{4-10}
    RevNet46 & Type I &  & 6.7M & 0.04 & 1.50x & 235 ($\uparrow$1.53x) & 0.85 & 1.01 & 1.85 \\	
    \cline{1-2}
    \cline{4-10}
    +SGD(8-bit) & -- &  & 6.7M & 0.039 & 1.54x & 236 (+1) & 0.87 & 1.04 & 1.88 \\
    \cline{1-2}
    \cline{4-10}
    RevNet57 & Type II &  & 6.1M & 0.03 & 2.00x & 300 ($\uparrow$1.95x) & 0.89 & 0.98 & 1.83 \\
    \cline{1-2}
    \cline{4-10}
    +SGD(8-bit) & -- &  & 6.1M & 0.029 & \textbf{2.07x} & \textbf{301} (+1) & 0.92 & 1.03 & 1.87 \\
    \hline
    \hline
    
    ResNet101  & -- & Bottleneck & 15.9M & 0.33 & -- & 31 & 0.62 & 0.80 & 1.48 \\
    \hline
    RevNet126 & Type I & \multirow{4}{*}{Basic} & 15.0M & 0.04 & 8.25x & 213 ($\uparrow$6.87x) & 0.58 & 0.80 & 1.48 \\	
    \cline{1-2}
    \cline{4-10}
    +SGD(8-bit) & -- &  & 15.0M & 0.039 & 8.46x & 215 (+2) & 0.62 & 0.83 & 1.49 \\
    \cline{1-2}
    \cline{4-10}
    RevNet137 & Type II &  & 14.2M & 0.03 & 11.00x & 297 ($\uparrow$9.58x) & 0.58 & 0.80 & 1.49 \\
    \cline{1-2}
    \cline{4-10}
    +SGD(8-bit) & -- &  & 14.2M & 0.029 & \textbf{11.37x} & \textbf{299} (+2) & 0.62 & 0.83 & 1.51 \\
    \hline
    RevNet140 & Type I & \multirow{4}{*}{Bottleneck} & 15.8M & 0.15 & 2.20x & 62 ($\uparrow$2.00x) & 0.60 & 0.80 & 1.43 \\	
    \cline{1-2}
    \cline{4-10}
    +SGD(8-bit) & -- &  & 15.8M & 0.147 & 2.24x & 63 (+1) & 0.62 & 0.82& 1.47 \\
    \cline{1-2}
    \cline{4-10}
    RevNet155 & Type II &  & 15.6M & 0.12 & 2.75x & 74 ($\uparrow$2.39x) & 0.59 & 0.81 & 1.45 \\
    \cline{1-2}
    \cline{4-10}
    +SGD(8-bit) & -- &  & 15.6M & 0.118 & 2.80x & 75 (+1) & 0.62 & 0.83 & 1.49 \\
    \hline
    \hline

    ResNet152  & -- & Bottleneck & 19.8M & 0.47 & -- & 22 & 0.55 & 0.74 & 1.39 \\
    \hline
    RevNet178 & Type I & \multirow{4}{*}{Basic} & 18.3M & 0.04 & 11.75x & 211 ($\uparrow$9.59x) & 0.54 & 0.75 & 1.41 \\	
    \cline{1-2}
    \cline{4-10}
    +SGD(8-bit) & -- &  & 18.3M & 0.039 & 12.05x & 213 (+2) & 0.57 & 0.77 & 1.43 \\
    \cline{1-2}
    \cline{4-10}
    RevNet197 & Type II &  & 18.2M & 0.03 & 15.67x & 295 ($\uparrow$13.41x) & 0.53 & 0.76 & 1.42 \\
    \cline{1-2}
    \cline{4-10}
    +SGD(8-bit) & -- &  & 18.2M & 0.029 & \textbf{16.21x} & \textbf{297} (+2) & 0.55 & 0.76 & 1.44 \\
    \hline
    RevNet230 & Type I & \multirow{4}{*}{Bottleneck} & 19.6M & 0.20 & 2.35x & 48 ($\uparrow$2.18x) & 0.49 & 0.72 & 1.33 \\	
    \cline{1-2}
    \cline{4-10}
    +SGD(8-bit) & -- &  & 19.6M & 0.195 & 2.41x & 49 (+1) & 0.52 & 0.73 & 1.36 \\
    \cline{1-2}
    \cline{4-10}
    RevNet245 & Type II &  & 19.4M & 0.12 & 3.92x & 73 ($\uparrow$3.32x) & 0.50 & 0.73 & 1.35 \\
    \cline{1-2}
    \cline{4-10}
    +SGD(8-bit) & -- &  & 19.4M & 0.118 & 3.98x & 74 (+1) & 0.53 & 0.75 & 1.38 \\
    \hline
  \end{tabular}
  \end{adjustbox}
\end{table*}

\subsection{System Description}
In this paper, we utilize ResNets and DF-ResNets as speaker embedding extractors, which are widely adopted networks and deliver state-of-the-art performance. The specific details as outlined below:
\begin{itemize}
    \item ResNet~\cite{rvector}: This is a popular and robust model in the speaker verification field. We use ResNet34, ResNet101, and ResNet152 as the baseline systems in our experiments.
    \item DF-ResNet~\cite{df-resnet}: It is an improved version of ResNet. We include DF-ResNet56, DF-ResNet110, DF-ResNet179, and DF-ResNet233 as base models.
\end{itemize}

\subsection{Training Settings}
During training, we randomly select a 200-frame chunk from each training speech sample and extract 80-dimensional FBANK features as input, using a 25-millisecond window length and a 10-millisecond frame shift. The loss function employs AAM-Softmax~\cite{aam}, with a margin of 0.2 and a scale of 32. We set the speaker embedding dimension to 256. For the ResNet baseline systems, we apply the SGD optimizer, setting the momentum to 0.9 and the weight decay to $1e^{-4}$. For DF-ResNet models, the AdamW optimizer is utilized with a weight decay of 0.05. The same training configuration is adopted for both the RevNets and DF-RevNets. In the optimizer quantization experiments, 8-bit versions of the SGD and AdamW optimizers are used.

\subsection{Evaluation Metrics}
For evaluation, the similarity among speaker embeddings is computed using cosine distance. Furthermore, we apply adaptive score normalization (AS-Norm)~\cite{asnorm} to calibrate score distribution, utilizing an imposter cohort size of 600. The system performance is assessed with the equal error rate (EER).

\section{Results and Analysis}
This section evaluates and analyzes the results of the reversible neural network and optimizer dynamic quantization algorithm introduced in Section \ref{sec:revnets} and \ref{sec:optimizer_quant}. First, Section \ref{ssec:revnet_results} assesses the performance of the two types of reversible neural networks based on ResNets and DF-ResNets. Subsequently, Section \ref{ssec:quant_results} presents the outcomes of the 8-bit dynamic quantization algorithm applied to the optimizers. Finally, Section \ref{ssec:mem_analysis} provides a detailed analysis of the relationship between GPU memory usage and  the number of parameters.

\subsection{Evaluation on Reversible Neural Networks}
\label{ssec:revnet_results}
We firstly evaluate the performance of the Type I and Type II reversible neural networks designed for ResNets and DF-ResNets, as introduced in Section \ref{sec:revnets}. The detailed results are presented below.

\subsubsection{Model Configuration}
In the experiments, we use the original ResNets (ResNet34, 101, and 152) and DF-ResNets (DF-ResNet56, 110, 179, and 233) as baseline systems. As described in Section \ref{sec:revnets}, to ensure a fair comparison, we design two types of reversible neural networks for each baseline model while keeping the number of parameters similar. For example, as shown in Table \ref{tab:revnets}, we introduce the Type I reversible neural network RevNet46 and the Type II reversible neural network RevNet57 based on the \textit{Basic} block, both of which have parameter counts similar to ResNet34. Similarly, four different reversible architectures are developed for ResNet101 and ResNet152, with comparable or fewer parameters. For DF-ResNets, we develop two reversible architectures for each baseline model, ensuring their parameters are comparable or lower, as Table \ref{tab:dfrevnets} illustrates.

\subsubsection{Performance Comparison}
From Table \ref{tab:revnets}, it is evident that our proposed reversible architectures exhibit comparable performance to the original ResNets across various parameter ranges (6M $\sim$ 20M). Specifically, RevNet46 and RevNet57 achieve similar EER to ResNet34 on the Vox1-O, E, and H. Interestingly, reversible neural network variants for ResNet101 and 152 can be constructed using the \textit{Basic} block. For example, RevNet126 and RevNet137, both based on the \textit{Basic} block, deliver performance comparable to ResNet101, which utilizes the \textit{Bottleneck} block. This highlights the strong representational capability of the \textit{Basic} block in reversible networks. Additionally, compared to RevNets built with the \textit{Bottleneck} block, those based on the \textit{Basic} block can achieve similar performance with fewer parameters. A similar trend is observed in DF-ResNets. As shown in Table \ref{tab:dfrevnets}, two types of reversible versions demonstrate comparable performance to the original baseline models while using equivalent or fewer parameters. These findings reveal the potential of reversible neural networks in maintaining strong speaker representation capabilities, confirming the effectiveness of the reversible operator.

\subsubsection{Memory Savings}
To evaluate the memory consumption of each model, we measure the memory usage and maximum batch size using a 2-second audio clip on a single 11 GB 2080Ti GPU. From Table \ref{tab:revnets}, we can see that the two proposed reversible networks significantly reduce memory usage across various configurations. For example, the reversible versions of ResNet34, namely RevNet46 and RevNet57, reduce memory usage by approximately half. Similarly, the reversible variants of ResNet101 and ResNet152, based on both \textit{Basic} and \textit{Bottleneck} architectures, also show significant memory reductions. Notably, the RevNets built on the \textit{Basic} block achieve even greater savings, with a maximum memory reduction of 15.7 times. This reduction stems from the ability of reversible networks to avoid storing intermediate activations during back-propagation. In addition, Type II reversible networks perform better than Type I in memory savings, as Type II replaces the irreversible downsampling layer with a reversible operation. Similarly, for DF-ResNets, both reversible neural network types achieve significant reductions in memory usage (Table \ref{tab:dfrevnets}). For example, DF-RevNet66 and DF-RevNet89 reduce GPU memory consumption by 4.6 times compared to the original DF-ResNet56 model. Under other model configurations, the maximum GPU memory compression is 14.2 times.

\begin{table*}[t]
  \caption{Architectural details and EER (\%) results of DF-ResNets and our proposed DF-RevNets on the Voxceleb1 dataset. The experimental settings are the same as ResNets and RevNets in Table \ref{tab:revnets}.}
  \label{tab:dfrevnets}
  \centering
  \setlength{\doublerulesep}{4.5pt}
  \begin{adjustbox}{width=.98\textwidth,center}
  \begin{tabular}{|l|c|c|c|c|c|c|c|c|c|}
    \hline
    \multirow{2}{*}{\textbf{System}} & \multirow{2}{*}{\textbf{Reversible}} & \multirow{2}{*}{\textbf{Block}} & \multirow{2}{*}{\textbf{\# Params}} & \multirow{2}{*}{\textbf{Memory (GB/utter)}} & \multirow{2}{*}{\textbf{Compression Ratio}} & \multirow{2}{*}{\textbf{Maximum Batch Size}} & \multirow{2}{*}{\textbf{Vox1-O}} & \multirow{2}{*}{\textbf{Vox1-E}} & \multirow{2}{*}{\textbf{Vox1-H}} \\ 
    & & & & & & & & &     \\
    \hline
    \hline
    DF-ResNet56 & -- & \multirow{5}{*}{DF-Bottleneck}  & 4.5M & 0.355 & -- & 33 & 0.96 & 1.09 & 1.99 \\
    \cline{1-2}
    \cline{4-10}
    DF-RevNet66 & Type I &  & 4.4M & 0.080 & 4.44x & 137 ($\uparrow$4.15x) & 0.84 & 1.07 & 1.98 \\	
    \cline{1-2}
    \cline{4-10}
    +AdamW(8-bit) & -- &  & 4.4M & 0.079 & 4.49x & 138 (+1) & 0.89 & 1.10 & 2.01 \\ 
    \cline{1-2}
    \cline{4-10}
    DF-RevNet89 & Type II &  & 4.1M & 0.077 & 4.61x & 141 ($\uparrow$4.27x) & 0.87 & 1.05 & 1.96 \\
    \cline{1-2}
    \cline{4-10}
    +AdamW(8-bit) & -- &  & 4.1M & 0.076 & \textbf{4.67x} & \textbf{142} (+1) & 0.90 & 1.07 & 1.98 \\
    \hline
    \hline
    
    DF-ResNet110 & -- & \multirow{5}{*}{DF-Bottleneck}  & 7.0M & 0.523 & -- & 23 & 0.75 & 0.88 & 1.64 \\
    \cline{1-2}
    \cline{4-10}
    DF-RevNet126 & Type I &  & 6.8M & 0.083 & 6.30x & 133 ($\uparrow$5.78x) & 0.71 & 0.87 & 1.65 \\	
    \cline{1-2}
    \cline{4-10}
    +AdamW(8-bit) & -- &  & 6.8M & 0.082 & 6.38x & 134 (+1) & 0.73 & 0.89 & 1.64 \\ 
    \cline{1-2}
    \cline{4-10}
    DF-RevNet149 & Type II &  & 6.5M & 0.077 & 6.79x & 137 ($\uparrow$5.96x) & 0.73 & 0.85 & 1.62 \\
    \cline{1-2}
    \cline{4-10}
    +AdamW(8-bit) & -- &  & 6.5M & 0.076 & \textbf{6.88x} & \textbf{138} (+1) & 0.75 & 0.87 & 1.65 \\ 
    \hline
    \hline

    DF-ResNet179 & -- & \multirow{5}{*}{DF-Bottleneck}  & 9.8M & 0.858 & -- & 14 & 0.62 & 0.80 & 1.51 \\
    \cline{1-2}
    \cline{4-10}
    DF-RevNet258 & Type I &  & 9.5M & 0.075 & 11.44x & 137 ($\uparrow$9.79x) & 0.59 & 0.79 & 1.50 \\	
    \cline{1-2}
    \cline{4-10}
    +AdamW(8-bit) & -- &  & 9.5M & 0.074 & 11.59x & 138 (+1) & 0.61 & 0.82 & 1.53 \\ 
    \cline{1-2}
    \cline{4-10}
    DF-RevNet281 & Type II &  & 9.1M & 0.073 & 11.75x & 140 ($\uparrow$10.00x) & 0.62 & 0.81 & 1.53 \\
    \cline{1-2}
    \cline{4-10}
    +AdamW(8-bit) & -- &  & 9.1M & 0.072 & \textbf{11.92x} & \textbf{141} (+1) & 0.61 & 0.83 & 1.52 \\ 
    \hline
    \hline

    DF-ResNet233 & -- & \multirow{5}{*}{DF-Bottleneck}  & 12.3M & 1.034 & -- & 12 & 0.58 & 0.76 & 1.44 \\
    \cline{1-2}
    \cline{4-10}
    DF-RevNet354 & Type I &  & 11.9M & 0.075 & 13.78x & 146 ($\uparrow$12.17x) & 0.59 & 0.75 & 1.45 \\	
    \cline{1-2}
    \cline{4-10}
    +AdamW(8-bit) & -- &  & 11.9M & 0.074 & 13.97x & 147 (+1) & 0.60 & 0.77 & 1.47 \\ 
    \cline{1-2}
    \cline{4-10}
    DF-RevNet377 & Type II &  & 11.5M & 0.073 & 14.16x & 149 ($\uparrow$12.42x) & 0.60 & 0.77 & 1.46 \\
    \cline{1-2}
    \cline{4-10}
    +AdamW(8-bit) & -- &  & 11.5M & 0.072 & \textbf{14.36x} & \textbf{150} (+1) & 0.59 & 0.79 & 1.45 \\ 
    \hline
  \end{tabular}
  \end{adjustbox}
\end{table*}

\subsubsection{Maximum Batch Size}
To determine the maximum batch size, our goal is to find the largest number of speech samples that can fit into each training batch without exceeding GPU memory limits. The batch size is inversely related to the memory usage per sample, meaning that the more memory saved, the larger the batch size that can be accommodated. For example, RevNet57 allows for a 1.9x increase in batch size compared to ResNet34. Likewise, RevNet197, a reversible variant of ResNet152, increases the maximum batch size from 22 to 295 samples. In the case of DF-ResNet233, the batch size expands from 12 to 149. These results highlight the potential for training deeper models on consumer GPUs.

\subsection{Evaluation on Optimizer State Quantization}
\label{ssec:quant_results}
In this part, we provide the results of the 8-bit dynamic quantization algorithm for optimizer states proposed in Section \ref{sec:optimizer_quant}. The details are presented below.

We first introduce the experimental setups. ResNets and RevNets are trained using the SGD optimizer with momentum, while DF-ResNets and DF-RevNets are trained with the AdamW optimizer. As described in Section \ref{sec:optimizer_quant}, the proposed dynamic quantization algorithm is applied to both SGD and AdamW, yielding their respective 8-bit versions. To evaluate the performance of these optimizers, they are directly adopted during the training of reversible neural networks. Specifically, RevNets are trained using the 8-bit version of SGD, while DF-RevNets are trained with the 8-bit version of AdamW.

Table \ref{tab:revnets} and \ref{tab:dfrevnets} clearly demonstrate that the 8-bit version of the optimizer reduces memory usage during training while maintaining model performance. For example, in RevNet46 and RevNet57, replacing the original 32-bit SGD optimizer with its 8-bit variant decreases memory usage per training sample from 0.04GB to 0.039GB and from 0.03GB to 0.029GB, respectively, and increases the maximum batch size by 1 with no performance loss. Similar results are observed for ResNet101 and ResNet152, where the 8-bit SGD optimizer improves the memory compression ratio of RevNet137 from 11 to 11.3 and increases the maximum batch size by 2. Likewise, in RevNet197, the compression ratio improves from 15.7 to 16.2, and the batch size increases by 2 without model performance loss.

Similar conclusions can be drawn from the 8-bit AdamW optimizer applied to DF-RevNets. For example, the maximum batch size for DF-RevNet66 increases by 1, and the memory compression ratio rises from 4.44 to 4.49. In DF-RevNet377, the compression ratio increases from 14.2 to 14.4, with no impact on performance. These results confirm that the 8-bit optimizer performs on par with its 32-bit counterpart and demonstrate the effectiveness of the dynamic quantization algorithm.

\subsection{Memory Usage Analysis}
\label{ssec:mem_analysis}
In this part, we present a detailed analysis of the relationship between GPU memory usage and the number of parameters for the original ResNets, DF-ResNets, and their corresponding reversible neural networks, RevNets and DF-RevNets.

\begin{figure*}[t]
  \centering
  \includegraphics[width=\linewidth]{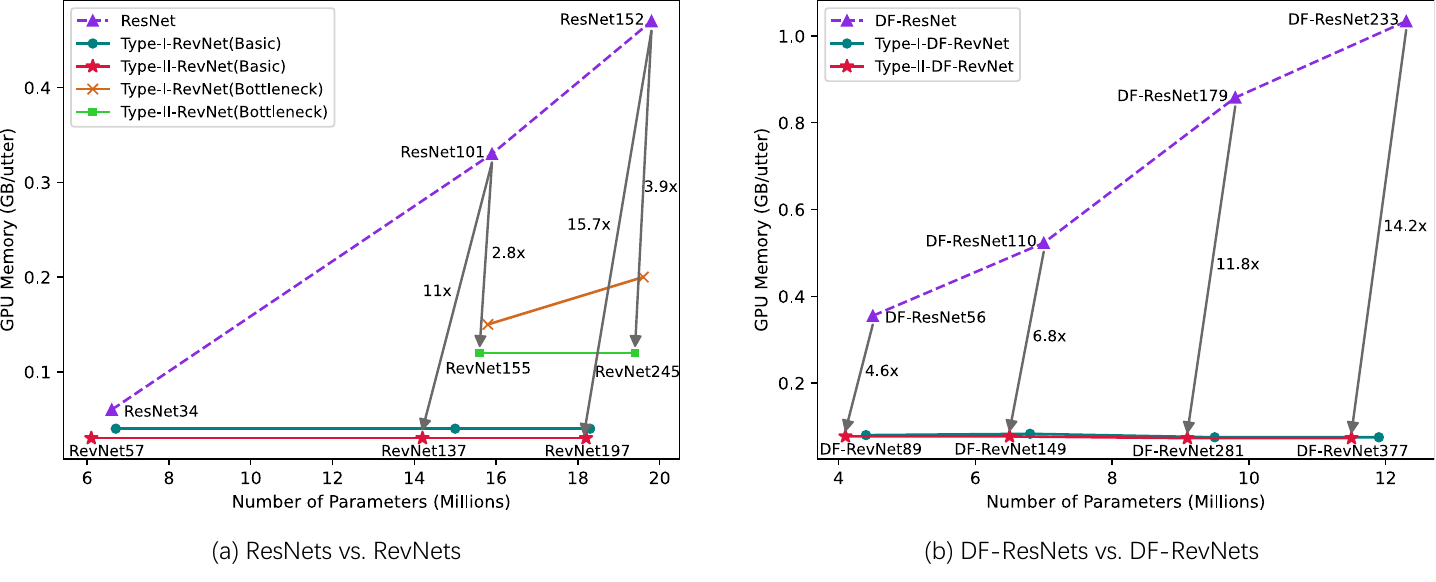}
  \caption{GPU Memory Usage vs. Parameter Number for ResNets and DF-ResNets.}
  \label{fig:mem_params}
\end{figure*}

\begin{figure*}[t]
  \centering
  \includegraphics[width=\linewidth]{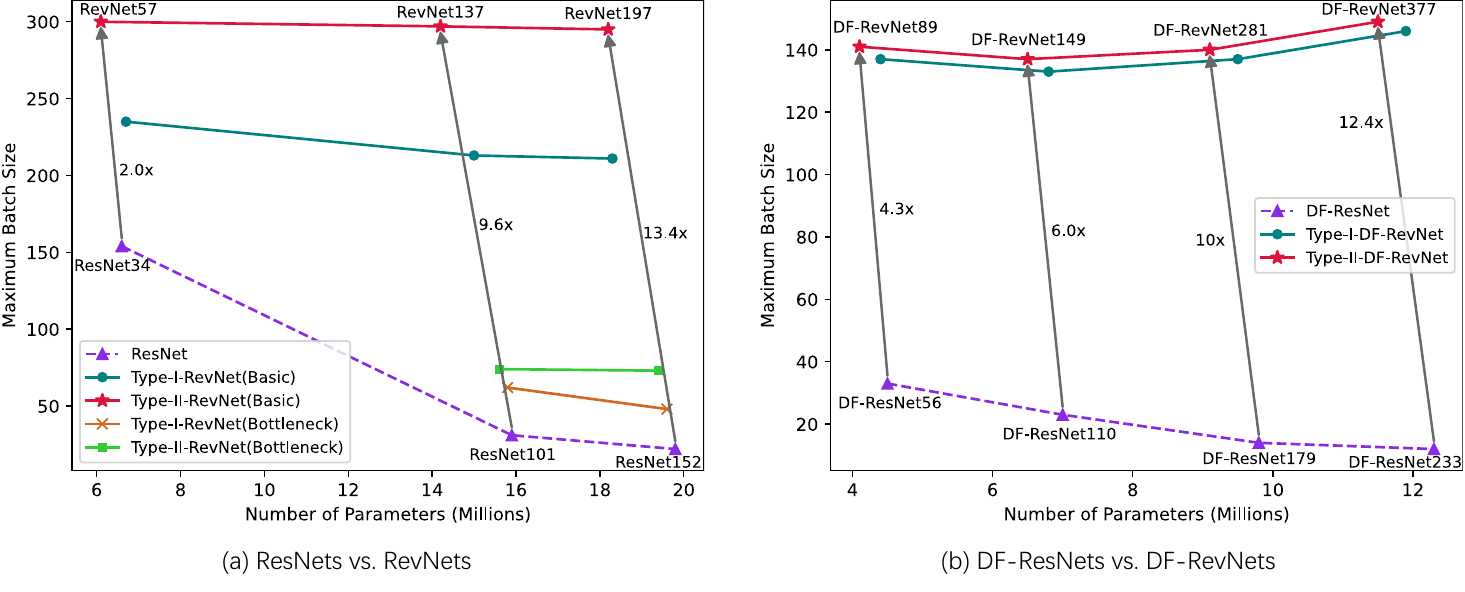}
  \caption{Maximum Batch Size vs. Parameter Number for ResNets and DF-ResNets.}
  \label{fig:batchsize_params}
\end{figure*}

\subsubsection{Memory vs. \# Params}
As shown in Fig. \ref{fig:mem_params}, the memory usage of the original ResNets and DF-ResNets increases linearly with network depth, which is consistent with the theoretical analysis of the back-propagation algorithm in Section \ref{ssec:back_prob}. This occurs because intermediate activations need to be cached during back-propagation, and their storage grows proportionally to the number of network layers. In contrast, both types of our RevNets and DF-RevNets maintain nearly constant memory consumption regardless of network depth. For example, Type-I RevNets and Type-II RevNets require a fixed memory of 0.04 GB and 0.03 GB per speech sample, respectively. This characteristic originates from the reversible nature of RevNets, which eliminates the need to store intermediate activations during back-propagation. As a result, the memory consumption is independent of network depth. In addition, Type-II reversible networks show superior GPU memory efficiency compared to Type-I. For example, Type-II RevNets achieve up to 40\% GPU memory savings over Type-I RevNets with similar parameters and performance. Our proposed RevNets achieve up to 15.7 times GPU memory savings while maintaining similar parameters and performance to the original ResNets. 

Similar phenomena can be seen with DF-ResNets. As Fig. \ref{fig:mem_params} displays, the memory usage grows linearly with the number of layers. For Type I and Type II DF-RevNets, GPU memory consumption remains constant at 0.08 GB and 0.073 GB, respectively, offering up to 14.2 times memory savings while retaining similar parameters and performance. The above analyses demonstrate the significant potential for achieving memory-efficient training in speaker verification systems.

\subsubsection{Maximum Batch Size vs. \# Params}
As illustrated in Fig. \ref{fig:batchsize_params}, the maximum batch size decreases as the number of parameters in the original ResNets and DF-ResNets grows. This is reasonable because larger models require more memory to store intermediate activations, and the optimizer state also consumes additional GPU memory. This observation further supports the GPU memory allocation analysis discussed in Section \ref{ssec:alloc_analysis}. Unlike the original models, reversible neural networks can maintain a stable maximum batch size even as the model depth increases. For example, the maximum batch size of RevNet57, 137, and 197 remains around 300. Notably, DF-RevNets even show an increasing trend; for example, the maximum batch size of DF-RevNet377 is 8 higher than that of DF-RevNet89. This is also attributed to the reversible architecture, which removes the need to cache intermediate activations.

Additionally, Type II reversible networks achieve a greater increase in maximum batch size compared to Type I reversible networks due to their higher degree of reversibility. Compared to the original models, RevNets and DF-RevNets can increase the maximum batch size by up to 13.4 times and 12.4 times, respectively, while maintaining comparable parameter numbers and performance. This provides the possibility of saving GPU resources during the training phase.

\subsubsection{GPU Resource Requirements}
In this part, we analyze the GPUs requirements, training time and performance of various speaker verification systems. As mentioned in Section \ref{sec:intro}, the total batch size can not be too small, otherwise it will negatively affect gradient estimation and model performance. The total batch size for training is calculated as follows:

\begin{small}
\begin{align}
  \text{Total Batch Size} &= \text{Maximum Batch Size / GPU} \times \text{\# GPU} \label{eq:totalbs}
\end{align}
\end{small}

Due to the limited GPU memory capacity, a common approach to increase the total batch size is to adopt a distributed training strategy by utilizing more GPUs. In the experiments, we target a commonly used total batch size of 256 and measure the number of GPUs required in terms of 11GB 2080Ti. As shown in Table \ref{tab:gpu_num}, our proposed methods significantly reduce the number of GPUs required for training, particularly with deeper neural networks, while maintaining nearly identical performance and comparable training time. For example, training the original ResNet101 requires 9 2080Ti, whereas our proposed reversible network RevNet137 only requires 1 GPU to complete the training, reducing resource consumption by a factor of 9 while maintaining similar training time and performance. For ResNet152, the GPU requirement is reduced from 12 to 1. Greater GPU savings can be achieved with DF-ResNets, where the number of GPUs required drops from 22 to 2 at most. This greatly reduces the demand for training resources and making it possible to effectively train deep speaker embedding extractors with limited resources.

Moreover, we provide a comparison of training resource requirements across different GPU series. As Table \ref{tab:gpu_type} indicates, normally training DF-ResNet233 requires 4 A100 (40GB) GPUs, while its reversible variant, DF-ResNet377, can be trained on only 2 2080Ti (11GB) GPUs with almost the same performance. Regarding training time, as the A100 is approximately 1.9x faster than 2080Ti, training is completed in 2.5 days. This demonstrates that our proposed method enables the effective training of deep speaker embedding extractors on consumer GPUs.


\begin{table}[t]
\caption{Taking the total batch size of 256 as an example, we use 11GB 2080Ti to measure the required number of GPUs, training time and EER (\%) performance of various speaker verification systems.}
\label{tab:gpu_num}
\centering
\begin{tabular}{c|c|c|c}
  \toprule
  \textbf{System} & \textbf{\# GPU} & \textbf{Training Time} & \textbf{Vox1-H} \\
  \midrule
  ResNet101 & 9 & $\sim$ 2 days & 1.48 \\
  RevNet137 & 1 & $\sim$ 3 days & 1.49 \\
  \midrule
  ResNet152 & 12 & $\sim$ 3 days & 1.39 \\
  RevNet197 & 1 & $\sim$ 5 days & 1.42 \\
  \midrule
  DF-ResNet179 & 19 & $\sim$ 3 days & 1.51 \\
  DF-RevNet281 & 2 & $\sim$ 5 days & 1.53 \\
  \midrule
  DF-ResNet233 & 22 & $\sim$ 4 days & 1.44 \\
  DF-RevNet377 & 2 & $\sim$ 6 days & 1.46 \\
  \bottomrule
\end{tabular}
\end{table}

\begin{table}[t]
\caption{The comparison of GPUs requirements and EER (\%) performance for training DF-ResNet233 and its reversible version.}
\label{tab:gpu_type}
\centering
\begin{tabular}{c|c|c|c}
  \toprule
  \textbf{System} & \textbf{GPU} & \textbf{Training Time} & \textbf{Vox1-H} \\
  \midrule
  DF-ResNet233 & 4 $\times$ A100 (40GB) & $\sim$ 2.5 days  &  1.44 \\
  DF-RevNet377 & 2 $\times$ 2080Ti (11GB) & $\sim$ 6 days & 1.46 \\
  \bottomrule
\end{tabular}
\end{table}

\section{Conclusion}
In this paper, we delve into the memory-efficient training strategies for deep speaker embedding learning in speaker verification under the resource-limited environments. We begin by thoroughly analyzing GPU memory usage during the training of speaker verification systems, revealing that activations and optimizer states are the main contributors to memory consumption. To tackle with activations, we introduce two types of reversible neural networks, which recompute intermediate activations during the backward pass, substantially reducing memory usage without impairing performance. For optimizer states, we propose a dynamic quantization technique that employs an adaptive tree-based 8-bit data format to replace standard 32-bit floating-point numbers. Experiments on VoxCeleb datasets indicate that both reversible ResNets and DF-ResNets are capable of training effectively without the need to store activations. Furthermore, the 8-bit variants of SGD and Adam optimizer reduce the memory usage by 75\%, while still maintaining comparable performance to their 32-bit equivalents. A comprehensive comparison of memory usage and performance reveals that our proposed models yield up to 16.2x memory savings, with similar parameters and performance to the original systems. 
We can effectively train deep speaker embedding extractors using just 1 to 2 consumer GPUs, such as the 2080Ti, compared to the previous requirement of dozens of GPUs or high-end GPU like the A100.

\bibliographystyle{IEEEtran}
\bibliography{mybib}


 





\end{document}